\documentclass[11pt,twoside]{article}

\usepackage{asp2014}

\aspSuppressVolSlug
\resetcounters

\begin{document}

\title{Large Language Models: New Opportunities for Access to Science}

\author{Jutta Schnabel$^1$ \textit{for the KM3NeT Collaboration}}
\affil{$^1$Erlangen Centre for Astroparticle Physics, Friedrich-Alexander Universität Erlangen-Nürnberg, Erlangen, Germany; \email{jutta.schnabel@fau.de}}

\paperauthor{Jutta Schnabel}{jutta.schnabel@fau.de}{0000-0003-1233-7738}{FAU Erlangen-Nürnberg}{ECAP}{Erlangen}{}{91052}{Germany}

\begin{abstract}
The adaptation of Large Language Models like ChatGPT for information retrieval from
scientific data, software and publications is offering new opportunities to simplify access to
and understanding of science for persons from all levels of expertise. They can become tools
to both enhance the usability of the open science environment we are building as well as help
to provide systematic insight to a long-built corpus of scientific publications.
The uptake of Retrieval Augmented Generation-enhanced chat applications
in the construction of the open science environment of the KM3NeT neutrino detectors serves as a focus point to explore and exemplify prospects for the wider application of Large Language Models
for our science.
\end{abstract}

\section{Potentials of Large Language Models for Open Science in KM3NeT}
Scientific information is conveyed primarily through journal publications, which limits the reuse of analysis methods and results mainly to those directly provided in graphs and tables. Although publications lately are increasingly accompanied by data sets and software, making data and workflows interoperable and systematizing multiple analysis results for common workflow integration remains a challenge in high-energy particle physics.
For the KM3NeT experiment, data from particle interactions and environmental information from the detectors' locations in the Mediterranean sea can serve as basis for a multitude of analyses in the areas of astrophysics and particle physics, physics beyond the standard model, or sea science. The KM3NeT collaboration, therefore, aims to provide data to the various communities through dedicated interfaces\footnote{to the Virtual Observatory through \href{https://vo.km3net.de}{vo.km3net.de} and in the Open Data Center \href{https://opendata.km3net.de}{opendata.km3net.de}}, and is enabling open science through a comprehensive Open Science System (OSS) that includes software and container repositories as well as an education platform \citep{schnabel2021}.

In order to develop the interfaces for FAIR (Findable, Accessible, Interoperable, Reusable) data sharing, legacy data from the ANTARES telescope \citep{deJong2023} will be integrated in the OSS to develop metadata standards and will be used to refine the requirements for workflow developments. Publications from ANTARES and KM3NeT serve as basis for systematization of analysis outcomes, and software and workflow documentation may be used to improve analysis code generation. Making information easily findable and more comprehensible through language processing is the aim of integrating Large Language Model (LLM) tools in the OSS environment.

A wide variety of LLMs are now available, either as general-purpose models or fine-tuned for specific domains \citep{naveed2024}. LLM performance can be further enhanced through optimization techniques such as prompt engineering or Retrieval Augmented Generation (RAG), where LLM prompts are enriched with information from reference databases. Additionally, LLMs can be integrated with external applications to tackle complex, multi-step tasks. The number of tools designed specifically for scientific research is rapidly growing, but their performance needs to be evaluated against the given information retrieval task, necessitating also the construction of dedicated evaluation data sets. In KM3NeT, the LLMTuner package is therefore developed to provide LLM tool solutions in the context of KM3NeT's OSS developments.

\section{The LLMTuner package}

The LLMTuner package\footnote{see LLMTuner at \href{https://github.com/YouSchnabel/llmtuner}{github.com/YouSchnabel/llmtuner}} offers a python-based environment enhancing the RAG capabilities provided by the AnythingLLM-software\footnote{see the AnythingLLM software: \href{https://github.com/Mintplex-Labs/anything-llm}{github.com/Mintplex-Labs/anything-llm} for the full package description} with additional functionalities for data retrieval and transformation, LLM tool performance evaluation and several user interface options.

\subsection{AnythingLLM integration}
The AnythingLLM stack is deployed for KM3NeT as a Docker-based server instance which provides all functionalities made available by the software, namely interfaces to a wide range of current LLMs, various vector database options and embedder models, and custom agents. The LLMTuner package communicates with the server via its Application Programming Interface (API), while manual manipulation of the software setup is still possible through the AnythingLLM Graphical User Interface. The LLM tool creation in AnythingLLM is facilitated by setting up dedicated \textit{workspaces} on the instance, which provide a chat to interface with a given LLM, include options for prompt engineering and embedding of reference documents from the vector database, and offer further tuning of retrieval parameters. This allows establishing multiple workspaces tuned for different use case requirements.

The main shortcomings of the AnythingLLM software in its use for KM3NeT are limited options to include information from protected and customized server endpoints, accessing e.g. internal documentation, it lacking the ability to easily handle a large number of documents and automated updating, as well as not including an environment to easily compare and benchmark the workspace performances. All of this motivates the development of the LLMTuner package.

\subsection{LLMTuner setup}

\paragraph{Information retrieval and storage}
The backbone of LLMTuner is formed by the separated InfoBasis package\footnote{see the InfoBasis package: \href{hhttps://github.com/YouSchnabel/infobasis}{github.com/YouSchnabel/infobasis}}, which creates a local storage and configuration folder, a sqlite database to manage all relevant processing information and a configuration file to store document repository endpoint URLs and their related authentication tokens or passwords. The database tables are created using json schema files\footnote{see \href{https://json-schema.org/}{json-schema.org}}, and can be customized and added as needed. Interfaces to custom API endpoints can be created by deriving a new \textit{Talker} class plugin from the \textit{Interface} class, adding a download function for the respective content and a crosswalk between the API structure and database tables like the document index table inventorizing all external resources. Data provenance is tracked by assigning a document identifier to each resource, which also ensures the correct mapping for documents used in RAG to their original URL. An \textit{OpenTalker} class is provided to retrieve HTML pages from websites without access restrictions.

\paragraph{Document embedding and workspace creation}
The \textit{Interface} class is also used to connect to the AnythingLLM API, and the \textit{AnyLLMBuilder} and \textit{AnyLLMChatter} classes are provided to manage workspace creation of and chatting with a workspace respectively. The \textit{AnyLLMBuilder} handles staging documents from the document index by downloading the content locally, with optional format conversions and preprocessing steps applicable before embedding the document content in the AnythingLLM vector database. A specific workspace is then created by transferring selected documents to the workspace, tuning criteria for document retrieval and adding prompt extensions for interaction with the selected LLM. Reference snippets and chat history are handled by the \textit{AnyLLMChatter} class, which is also employed to evaluate workspace performance.

\paragraph{Performance evaluation}
Evaluation in LLMTuner utilizes huggingface's evaluate package\footnote{see documentation at \href{https://huggingface.co/docs/evaluate/index}{huggingface.co/docs/evaluate/index}} in the \textit{Evaluator} class, which handles the full test setup from creation of test data sets with prompts and expected reference replies, the generation of workspace replies and application of selected evaluation methods for the comparison of replies with the references. Test setup and results are stored in the local database for wide-ranging comparisons of different workspaces and their configuration.

\subsection{Package use and development}
Examples for the basic configuration, creation of \textit{Talker} plugins, document processing, workspace creation and evaluation are provided as Jupyter notebooks. Using the LLMTuner package, chat replies and references can be easily integrated directly in python workflows and analysis if e.g. replies are requested in a specific output format. In order to provide a pre-configured LLMTuner setup and the related workspaces through a fixed interface to the user, LLMTuner includes a Flask-based option\footnote{web application framework Flask \href{https://flask.palletsprojects.com/en/stable/}{flask.palletsprojects.com}} to employ a server-based chat service, and website embedding for a given chat dialog. Although AnythingLLM itself also provides a client embedding option, its use is limited by the custom document processing in the LLMTuner workflow which necessitates mapping between the AnythingLLM's origin of the embedded document and its original URL when providing references for a given reply.\\
Future developments are planned to include an extension of the preprocessing options, high-level result displays of evaluation results, an extended frontend for chat-tool display, and containerized deployment using Docker.

\section{Applications in KM3NeT}

\paragraph{Internal documentation retrieval}
The development of dedicated workspaces for KM3NeT has been prioritized aiming for the highest impact on collaboration work, starting by providing an internal tool for information retrieval from the internal documentation and gitlab instance, and published scientific contributions like papers and conference proceedings. Here, findability of relevant document snippets constitutes the most relevant evaluation criterion, which primarily focuses on retrieval quality from the vector database. As this also constitutes one of the most fundamental functionalities of RAG, the approach allows to develop related test data sets and establish the basic soundness of the framework. Beyond this, the test data sets serve to establish a first benchmark for LLM selection and prompting comparison to tweak the workspace reply to the required applications output. The interface to this assistant will be provided on an internal webpage as it might contain private information, and also the selection of the LLM used in the related workspace will be subject to the requirements for privacy.

\paragraph{Analysis workflow assistant and data retrieval}
In order to facilitate the construction of a workspace that focuses on assisting the researcher to create scientific workflows and analysis code, similar document resources as in the above case are relevant, but the LLM capabilities differ, as a larger understanding of the intent and analysis of the process steps is required to formulate meaningful answers, and LLM models finetuned to the use in e.g. programming might be required. This use case will also widen the target community for the application, as this assistant is also relevant for non-KM3NeT researchers, making it necessary to consider how to well distinguish between internal and external sources and focusing on how to improve the KM3NeT OSS resources to enhance the chat tool output.

\paragraph{General access and multilingual education}
In the most wide-reaching application, a general audience will be considered that is non-expert, multilingual and to which the tools should offer assistance to more easily understand the fundamental concepts and functioning of the KM3NeT detectors and astroparticle physics. This tool can tremendously help to improve e.g masterclasses by the respective member institutes or be used independently in higher secondary education or by students. Requirements here are to provide a tool that is efficient in translation and abstracting, and therefore focuses more on the LLM capabilities and prompt engineering than in the above cases. 

In these three applications, KM3NeT aims to improve the OSS to open up fundamental science and enhance the open science paradigm to provide science for all.

\acknowledgements The developement of the LLMTuner package is supported by the FAU Emerging Talents Initiative.

\bibliography{P204}  


\end{document}